\documentclass[logo,11pt,a4paper]{ETHpaper}

\usepackage{amsmath,amssymb} 
\usepackage{graphicx,tikz}
\usepackage{subfigure}

\begin{document}
%
\title{A Tunable Mechanism for Identifying Trusted Nodes \\ in Large Scale Distributed Networks}

\author{Joydeep Chandra$^{1,2}$, Ingo Scholtes$^{1}$, Niloy Ganguly$^2$, Frank Schweitzer$^1$}
\authoralternative{Joydeep Chandra, Ingo Scholtes, Niloy Ganguly, Frank Schweitzer}
\address{
  $^1$ Chair of Systems Design, ETH Z\"{u}rich, CH-8032 Z\"{u}rich, Switzerland \\
  $^2$ Department of Computer Science \& Engineering,\\ Indian Institute of Technology, Kharagpur 721302, India \\
  \texttt{\{jchandra,ischoltes,fschweitzer\}@ethz.ch, niloy@cse.iitkgp.ernet.in}
}
\date{\today}
\maketitle

\begin{abstract}
In this paper, we propose a simple randomized protocol for identifying trusted nodes based on
 personalized trust in large scale distributed networks. The problem of identifying trusted nodes, based on personalized trust, in a large network setting  stems from the huge computation and message overhead involved in exhaustively calculating and propagating the trust estimates by the remote nodes. However, in any practical scenario, nodes generally communicate with a small subset of nodes and thus exhaustively estimating the trust of all the nodes can lead to huge resource consumption. In contrast, our mechanism can be tuned to locate a desired subset of trusted nodes, based on the allowable overhead, with respect to a particular user. The mechanism is based on a simple exchange of random walk messages and nodes counting the number of times they are being hit by random walkers of nodes in their neighborhood. Simulation results to analyze the effectiveness of the algorithm show that using the proposed algorithm, nodes identify the top trusted nodes in the network with a very high probability by exploring only around 45\% of the total nodes, and in turn generates nearly 90\% less overhead as compared to an exhaustive trust estimation mechanism, named TrustWebRank. Finally, we provide a measure of the global trustworthiness of a node; simulation results indicate that the measures generated using our mechanism differ by only around 0.6\% as compared to TrustWebRank. 
\end{abstract}

\newpage

\section{Introduction}
The Internet has witnessed a huge growth not only in the number of users but also in the volume of contents and resources being shared between them. In recent times, with the advent of peer-to-peer (p2p) based services, online social networks (OSN) as well as platforms that facilitate the sharing of user-generated content, the importance of user contributions has increased significantly. However, as the amount of these contributions is growing in the Internet community, their trustworthiness is becoming an important concern: In the context of file-sharing and content distribution systems, malicious participants that deliberately distribute malware, fake files or mislabeled content constitute a significant problem. At the same time, p2p based services frequently suffer from \emph{freeriding} users which exploit the resources provided by other participants while not contributing resources in return. It has thus been acknowledged that it is important to consider trust and reputation management schemes in the design of any distributed system that relies on user contributions \cite{Wierzbicki-2010}.

Typically, such trust and reputation management aims at globally estimating the trustworthiness of users based on a record of previous cooperative or defective behavior. In file-sharing scenarios, this typically refers to the quality of files previously provided to other users, while in settings prone to freeriders, the amount of previously shared resources may be considered for this purpose. There exist several examples for trust management mechanisms which approach this problem by estimating the trustworthiness of users by a global, typically scalar value \cite{massa-recsys07,matsuo-www09, zhao-icccn09, zhou-ipdps06, bernard-2010}. 

A major drawback of such approaches to trustworthiness is the fact, that in general trust is a personalized concept, i.e. the level of trust users may have in the behavior of a particular participant will differ across different sets of users or user communities. At the same time, differences between user preferences may lead to situations where the trust in the quality of other users' content may vary significantly across different participants \cite{matsuo-www09}.

Due to the importance of providing \emph{personalized notions of trust} in a variety of distributed computing scenarios, recently a number of works have addressed this question \cite{li-ithet10, liu-cikm09, walter-recsys09}. Typically, these models rely on a transitive notion of indirect trust, i.e. if a node A trusts node B and node B trusts node C, then it is assumed that node A also trusts node C to a certain extent. A major benefit of this approach is that, based on a very sparse network of direct trust relations, it allows to make statements about the \emph{indirect trust between users that have not interacted so far}.

Interesting approaches have been studied which intend to capture this personalized and indirect trust. A major hurdle in applying them in practical settings is the huge communication and computation overhead that is necessary to exhaustively compute the indirect trust between all possible pairs of nodes. A particular observation that serves as a motivation for our scheme is however that the exhaustive computation of all pairs of indirect trust is often unnecessary. Instead, there are numerous situations where it is sufficient to present users with a \emph{personalized set containing only a limited number of most trusted peers}. Examples, where such a limited set of trusted peers can be useful are social recommender systems, peer selection schemes in file-sharing scenarios, as well as topology management and adaptation mechanisms in p2p-based distributed services. Furthermore, since the required number of trusted interaction partners can vary depending on the actual application, any practical scheme to retrieve such a list should provide the designer of a distributed system with the possibility to make a trade-off between the required number of most trusted nodes and the communication and computation overhead entailed by the protocol.

Following this motivation, in this article we introduce and evaluate a simple heuristic mechanism which attempts to solve the open problem of estimating indirect trust in a more efficient way. In particular, as will be argued in section \ref{s:analysis}, we show that the proposed protocol captures precisely the same transitive notion of personalized indirect trust as the \emph{TrustWebRank} algorithm which has been proposed in \cite{walter-recsys09}. We further show, that our protocol effectively constitutes a very simple heuristic that allows to sample indirect trust values to a variable degree of precision and completeness which can be tuned to the needs and resources of a particular application. In its essence, the protocol is a random message passing scheme in which nodes bias message passing probabilities according to the level of direct trust they hold in their previous interaction partners. The proposed protocol, as well as its relation to the TrustWebRank algorithm will be described in more detail in sections \ref{s:trust-persona} and \ref{s:heuristic}. We then evaluate the correctness of the proposed scheme with respect to how well it is able to recover the indirect trust computed by the exhaustive and analytical approach taken by the TrustWebRank algorithm. We further evaluate the efficiency in terms of the reduction of messages passed compared to the original analytical approach. Simulation results based on scale-free as well as Erd\"{o}s-R\'{e}nyi trust topologies indicate that with as much as a 90\% reduction in terms of message overhead, the proposed sampling algorithm can produce almost equivalent results as the TrustWebRank mechanism in terms of the identification of the most trusted remote nodes for each user in the network. We also propose a measure of global trustworthiness of the users in section \ref{s:global} and compare the same for the users when the trust values are calculated using our mechanism and TrustWebRank. We finally conclude our paper with some interesting directions for future research.

\begin{table}
\centering
\begin{tabular}{|c|l|}
\hline
$T_{ij}$ &  Normalized direct trust of node $i$ in node $j$\\
\hline
$\bf{T}$ & Direct trust matrix whose each element is $t_{ij}$\\
\hline
$\beta$ & Damping factor used in TrustWebRank mechanism\\
\hline
$\gamma$ & Damping factor used in our proposed mechanism\\
\hline
$N(i)$ & Neighbor set of node $i$\\
\hline
$\bf{N}$ & Diagonal matrix representing the number of \\& random walkers sent by each node\\
\hline
$\bf{H}$ & Matrix, whose entry $h_{ij}$ represents the \\& number of hits from source $i$ to $j$\\
\hline
$\bf{\hat{H}}$ & Matrix obtained by normalizing each entry of $\bf{H}$ over each row\\
\hline 
$\bf{S}$ & Matrix, whose entry $s_{ij}$ represents the trust rating\\& of source $i$ in $j$ calculated using TrustWebRank\\
\hline
$\bf{\hat{S}}$ & Matrix obtained by normalizing each entry of $\bf{S}$ over each row\\
\hline
\end{tabular}
\caption{ \footnotesize List of symbols}
\label{tab:symbols}
\end{table}

\newpage

\section{TrustWebRank}\label{s:trust-persona}
In this section we discuss about TrustWebRank, an existing personalized trust estimation mechanism, and discuss about the motivation of our proposed algorithm. A summary of the list of symbols used in this description as well as in the remainder of the paper is shown in table \ref{tab:symbols}. 
\par In TrustWebRank, each node, $i$, maintains individual 
trust opinions about certain other 
users (represented by $T_{ij}\in (0,1]$) with which it has direct interaction or have similarity in the preferences. These users are termed as 
direct neighbors of the node and
the measured trust in these users as the \emph{direct trust}. The trust estimation for user $j$ that is not a direct neighbor of node $i$ is
done by considering all the paths used to reach user $j$ weighted by the direct trust values of each link used in the path. It is necessary to explore all the paths, as more the number of paths to a user $j$, the higher will be its trust.  
The trust value calculated using this mechanism is termed as the \emph{indirect trust} of node $i$ on user $j$ and is represented as $\tilde{T}_{ij}$.
The TrustWebRank mechanism is an exhaustive mechanism for estimating trust between every pair of nodes
and can also be implemented in a distributed network as proposed in \cite{carchiolo-sci10}. The indirect trust values of the nodes
obtained using the TrustWebRank mechanism can be represented analytically as follows.
\par If $\bf{S}$ is a stochastic matrix such that the elements $S_{ij}=\frac{T_{ij}}{\sum_{k\in N(i)} T_{ik}}$, then the indirect trust 
of node $i$ in a node $l$ is given as 
\begin{eqnarray}
\tilde{T}_{il}=S_{ij} + \beta\sum_{k\in N(i)}S_{ik}\tilde{T}_{kl}\quad\forall i,l \label{eq:walter-baseq}
\end{eqnarray}
where $\beta \in [0,1)$ is a damping factor that reduces the trust value of node $l$ with increasing distance from $i$. If $\bf{\tilde{T}}$ 
denotes the matrix representing the indirect trust values then $\bf{\tilde{T}}$ can be represented as
\begin{eqnarray} 
 \bf{\tilde{T}}=(\bf{I}-\beta\bf{S})^{-1}\bf{S},\label{eq:trust_walterEq}
\end{eqnarray}
where $\bf{I}$ is the identity matrix. These indirect trust values can be normalized to form a row stochastic matrix $\bf{\hat{S}}$
whose elements are represented as 
\begin{eqnarray}
\hat{S}_{ij}=\frac{\tilde{T}_{ij}}{\sum_{k\in N(i)}\tilde{T}_{ik}} \label{eq:norm-TWR}
\end{eqnarray}
However, for large graphs, performing these matrix operations require
huge computational overhead. A practical distributed implementation scheme for the TrustWebRank mechanism
has been proposed in \cite{carchiolo-sci10}, where they proposed an iterative method to derive the trust values of
the node pairs. In this approach, equation \ref{eq:walter-baseq} is formulated as
\begin{eqnarray}
 \tilde{T}^{m+1}_{il}=S_{ij} + \beta\sum_{k\in N(i)}S_{ik}\tilde{T}^{m}_{kl},
\end{eqnarray}
where $\tilde{T}^{m}_{kl}$ is computed at step $m$. Thus the algorithm is a flooding based algorithm that iteratively explores the 
trust values of the nodes with increasing steps $m$ finally leading to a convergent solution. This scheme is exhaustive with little 
control over the number of nodes
that are to be explored and hence also on the message overhead involved. {\em However from a designer's perspective a practical alternative would be to design a tunable algorithm that can provide the users an option to consider a trade-off between the required number of trusted 
nodes and the allowable message overhead}. Users willing to explore more number of trusted nodes will have to pay with a higher overhead.
This is suitable in realistic scenarios, e.g. during bootstrapping in p2p networks like Gnutella \cite{karbhari-pam04}, a newly arriving node need not 
evaluate the trust to all the nodes in the network. Rather it will preferably connect to a small number of trusted peers and hence need to
obtain the trust of only a small subset of the whole network.  Thus, from the perspective of an individual node, it would be more useful to 
identify certain number of 
most trusted peers with which it can communicate. In this paper, we exploit this concept to propose a tunable mechanism based on random walks that preferentially samples 
a set of nodes based on their trust values (evaluated using the TrustWebRank mechanism) with respect to
a user in a network. The parameters in the algorithm can be tuned to increase or decrease the number of trusted nodes that are explored based on the allowable message overhead.
The proposed mechanism efficiently selects trusted nodes with high probability besides being lightweight and scalable. We next detail our proposed mechanism.
\section{Proposed Random Walk Based Mechanism}\label{s:heuristic}
Similar to the TrustWebRank, we initially normalize the direct trust values to all the neighbors of a node to obtain the stochastic matrix $\bf{S}$ given as 
\begin{eqnarray}
 S_{ij}=\frac{T_{ij}}{\sum_{l\in N(i)}T_{il}}
\end{eqnarray}
The normalized trust values provide measure of the relative trust in the neighbors of each node. 
To derive the 
indirect trust of the remote nodes, each node initially sends $W$ random walkers, each traversing through a link with a probability equal to the normalized direct trust value of the node connected 
through that link. Thus for a node $i$,
a random walker visits a neighbor $j$ with probability $S_{ij}$. Hence for $W$ random walkers the average number of 
walkers that visit node $j$ from node $i$ is given as 
\begin{eqnarray}
 V_j^{(i)}=W\cdot S_{ij}.
\end{eqnarray}
When a walker
from a source node $i$ reaches a node $j$ on the $h^{th}$ hop, node $j$  stamps its ID on the walker message and does any of the following two steps.
\begin{enumerate}
 \item With a probability, $\gamma^{h}S_{jk}$, it forwards the message to a neighbor node $k$, where $\gamma$ is the damping factor, similar to $\beta$ that is used in TrustWebRank,  
with a value that lies between $(0, 1)$. 
\item With a probability $1-\sum_{l\in N(j)}(\gamma)^{h}S_{jl}$, the random walker dies, i.e. $j$ does not forward the message further, instead it sends back the message to the source node $i$. 
\end{enumerate}
Thus $\gamma$ controls the number of hops of the random walker and hence 
we do not use an explicit Time-to-Live (TTL) value of these walkers. The major importance of using this damping factor is that nodes which are nearer to a source will receive more number of hits as compared to distant nodes. The rationale behind this is, it is difficult to predict the trust value of a node which is
far away from a source. Thus the trust value of a node will
decrease with path length from the source node. 
\par On receiving the $W$ messages back, the source node calculates the number of times each node has been hit that provides a measure of
trust to the remote nodes. Thus the precision of the trust values depends on both the number of walkers as well as the damping factor $\gamma$ and hence need to be carefully tuned so as to control 
the message overhead besides maintaining a certain level of precision. We discuss the effects of both 
these parameters later in section \ref{s:analysis}. Thus we can represent the probability that a random walker that has 
reached a node $j$ on the $h^{th}$ hop, traverses the link to a neighbor node $k$ in the next hop by 
\begin{eqnarray}
 H^{(h)}_{jk}=\gamma^{h}T_{jk}
\end{eqnarray}
and the probability that the random walker dies at node $j$ in the $h^{th}$ step by, 
\begin{eqnarray}
 D^{(h)}_{j}=1-\sum_{l\in N(j)}H^{(h)}_{jl}=1-\sum_{l\in N(j)}\gamma^{h}T_{jl},
\end{eqnarray}
Thus the average indirect trust values of each pair of peers for our proposed algorithm can be derived as follows. 
The stochastic matrix $\bf{S}$ provides the probability that a random walker from node $i$ will
reach a neighbor $j$ in one step. The damping factor $\gamma$ reduces the probability of a walker to continue further
after each hop; thus the matrix representing the probabilities that a random walker from any node $i$ reaches node $j$ after exactly
$l$ hops (although it might have reached $j$ in earlier hops) is given as $\gamma^{l-1}\bf{S}^{\it{l}}$. So the average number of times node
$j$ has been hit from source $i$ by a single random walker can be written using a matrix notation as 
\begin{eqnarray}
 \bf{H_1}&=&\frac{1}{\gamma}\left[(\gamma{\bf S})^1+(\gamma{\bf S})^2+\ldots \infty\right]\nonumber\\
&=& \left(\bf{I}-\gamma\bf{S}\right)^{-1}\bf{S},\label{eq:trust_myEq}
\end{eqnarray}
where $\bf{I}$ is the identity matrix. Note, equation \ref{eq:trust_myEq} provides the indirect
trust between any pair of nodes and is exactly similar to equation \ref{eq:trust_walterEq} of the 
TrustWebRank scheme. Thus this proves that our random walk based method produces consistent trust
estimates
with the TrustWebRank mechanism. To implement this mechanism, we send a number of random walkers
from each node and if $\bf{N}$ be a diagonal matrix that represents the
number of random walkers sent by each node, then, the average number of hits at each node for every
source node can be represented as $\bf{H}=\bf{NH_1}$. 
\par As the trust in a node $j$ by node $i$ is estimated 
by the number of hits to node $j$ of the random walkers from $i$, one should note that node $j$ should be hit at least once from $i$
to have a measure of the trust on it. Thus to ensure that at least one random walker reaches to all other nodes in the connected component, each node will have to theoretically send infinitely large 
number of random walkers. Hence, by sending infinitely large number of random walkers from each node, the nodes can obtain the same relative trust 
value in the other nodes as is measured using the TrustWebRank. However, it is obvious that sending such a large
number of random walkers is infeasible due to the huge message overhead it would generate. Hence, the tunability of algorithm is achieved
by controlling the number of random walkers, where a high number of random walkers will help to 
explore a large subset of nodes in order of their trust but at the expense of a higher message overhead.
\par We can normalize each row of the matrix, $\bf{H}$, to produce a row stochastic matrix, $\bf{\hat{H}}$, where
each  element $\hat{h}_{ij}$ will represent the relative trust of node $i$ on $j$ over all the nodes in the network. We use the matrix, $\bf{\hat{H}}$, along with $\bf{\hat{S}}$, the normalized trust matrix obtained in the TrustWebRank mechanism, to analyze the preciseness and efficiency of our proposed algorithm.
We perform a comparative study in the next section based on the following issues that we need to analyze:
\begin{enumerate}
\item How does the trust values correlate with the trust values calculated using TrustWebRank for various kind of topologies?
\item How does the number of random walkers and the damping factor influence the preciseness of the trust values?
\item How much performance efficiency is achieved by using our algorithm as compared to the TrustWebRank?
\end{enumerate}
We seek answers to these questions in the next section.

\begin{figure*}[!htbp]
\centering
\subfigure[]{\includegraphics[width=2in,scale=1.0]{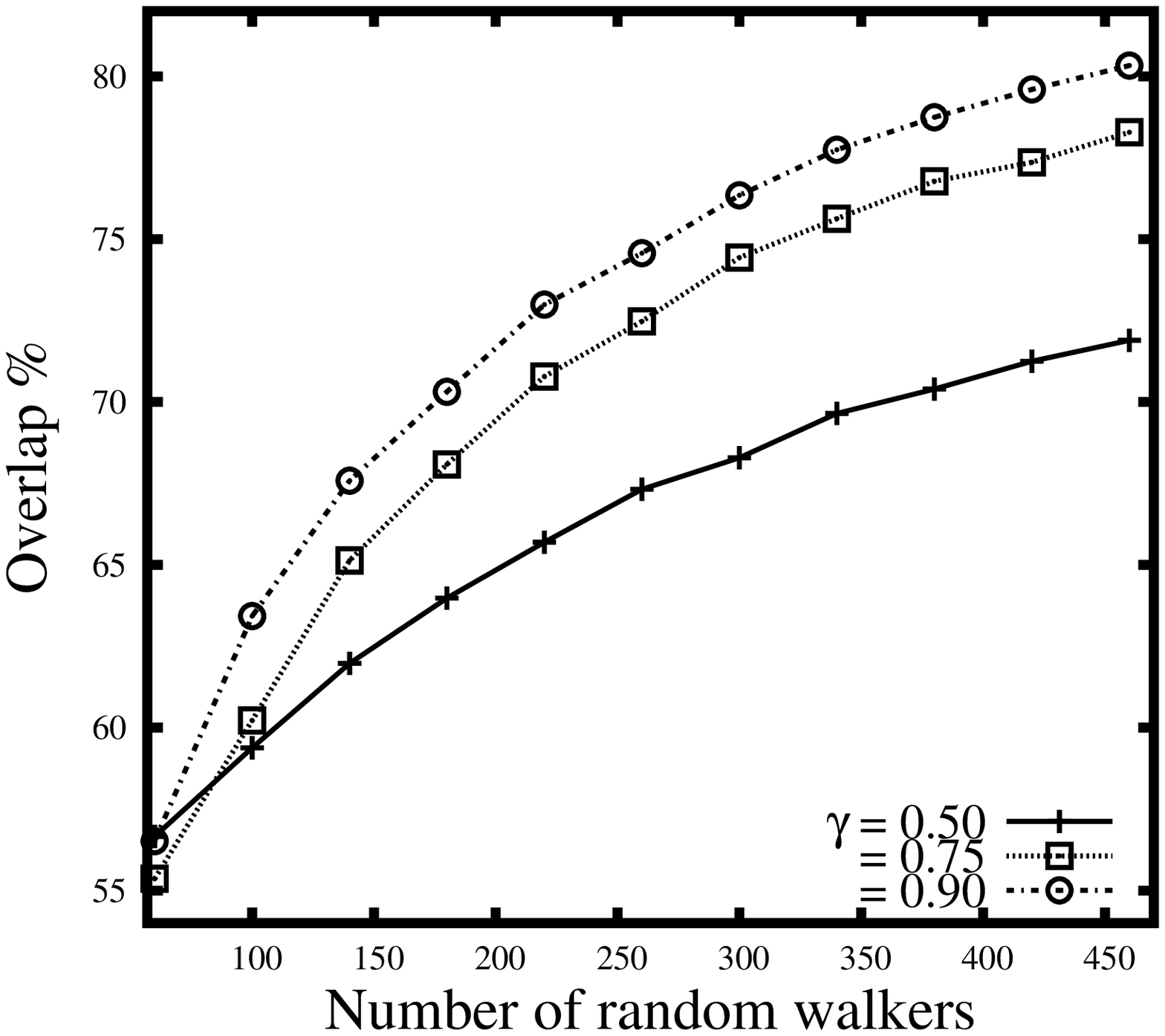}}
\subfigure[]{\includegraphics[width=2in,scale=1.0]{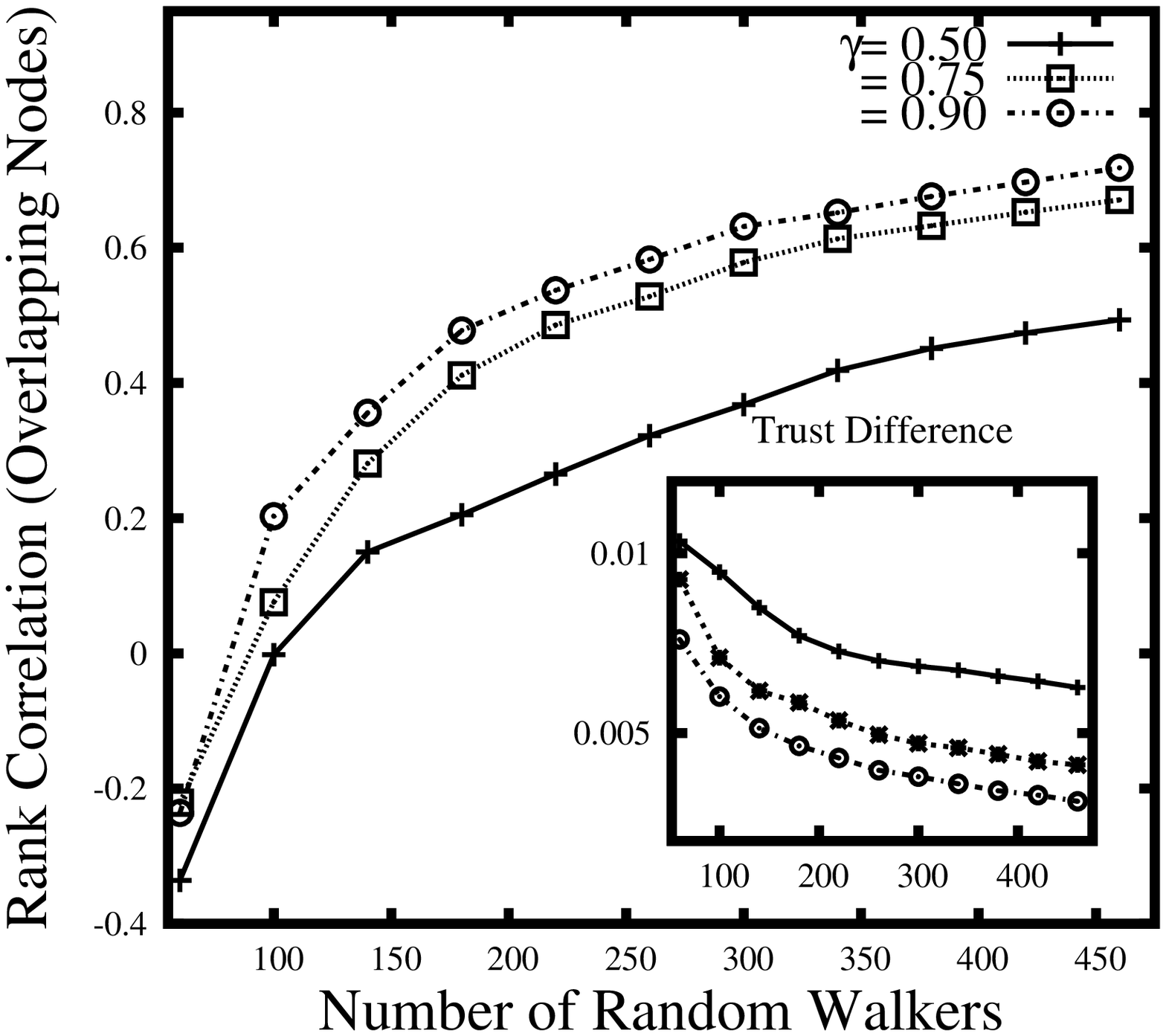}}
\subfigure[]{\includegraphics[width=2in,scale=1.0]{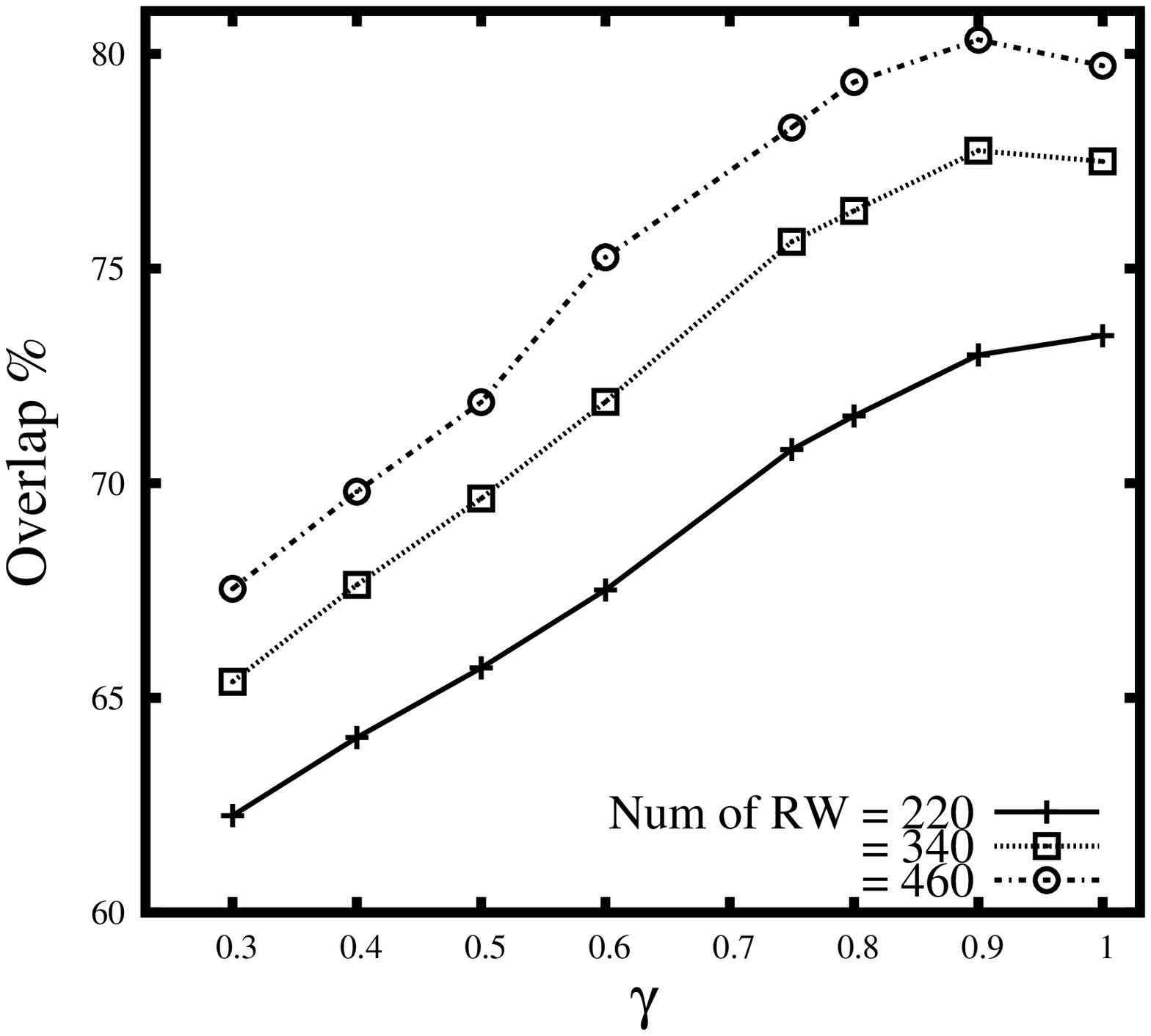}}
\subfigure[]{\includegraphics[width=2in,scale=1.0]{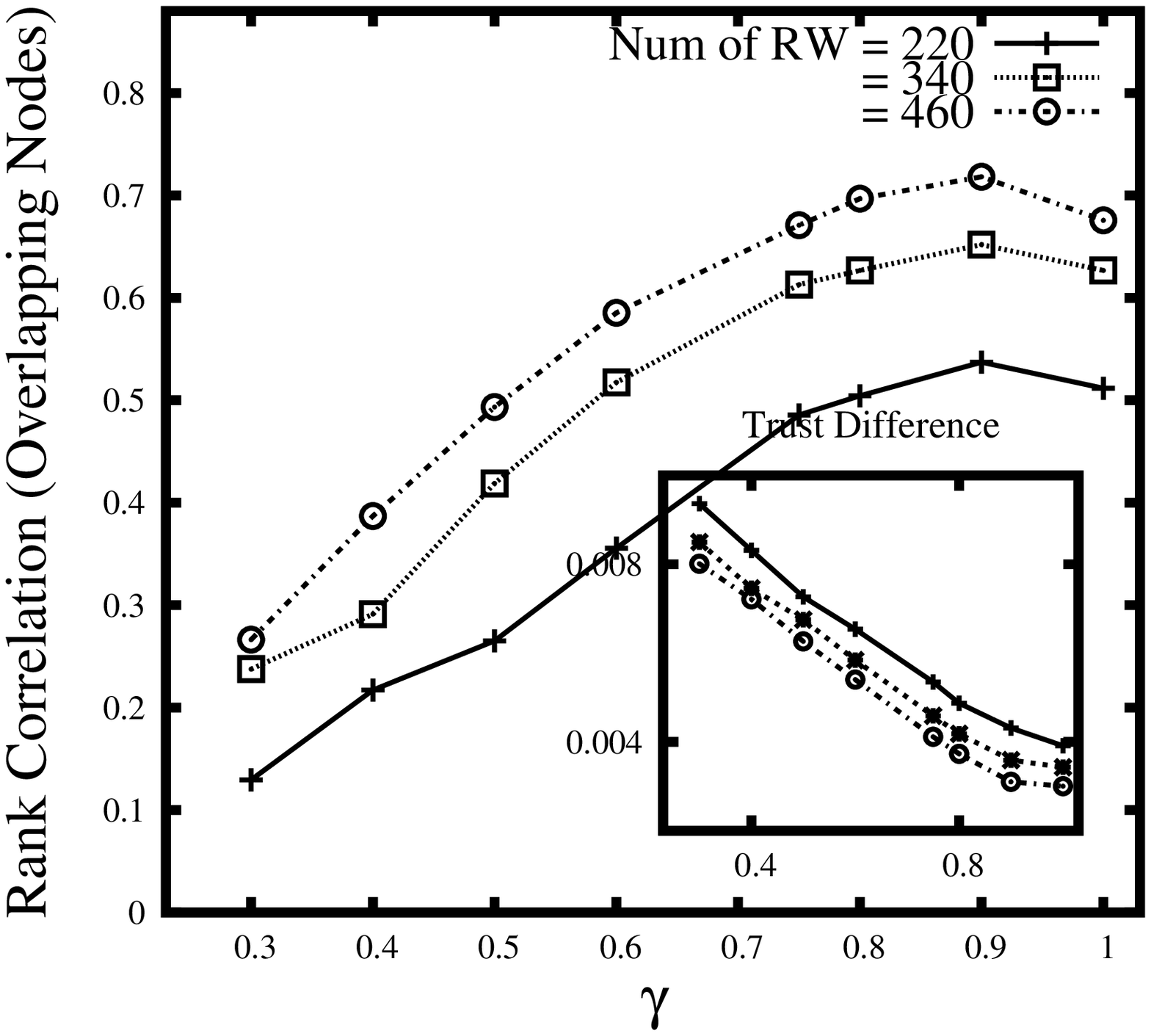}}
\subfigure[]{\includegraphics[width=2in,scale=1.0]{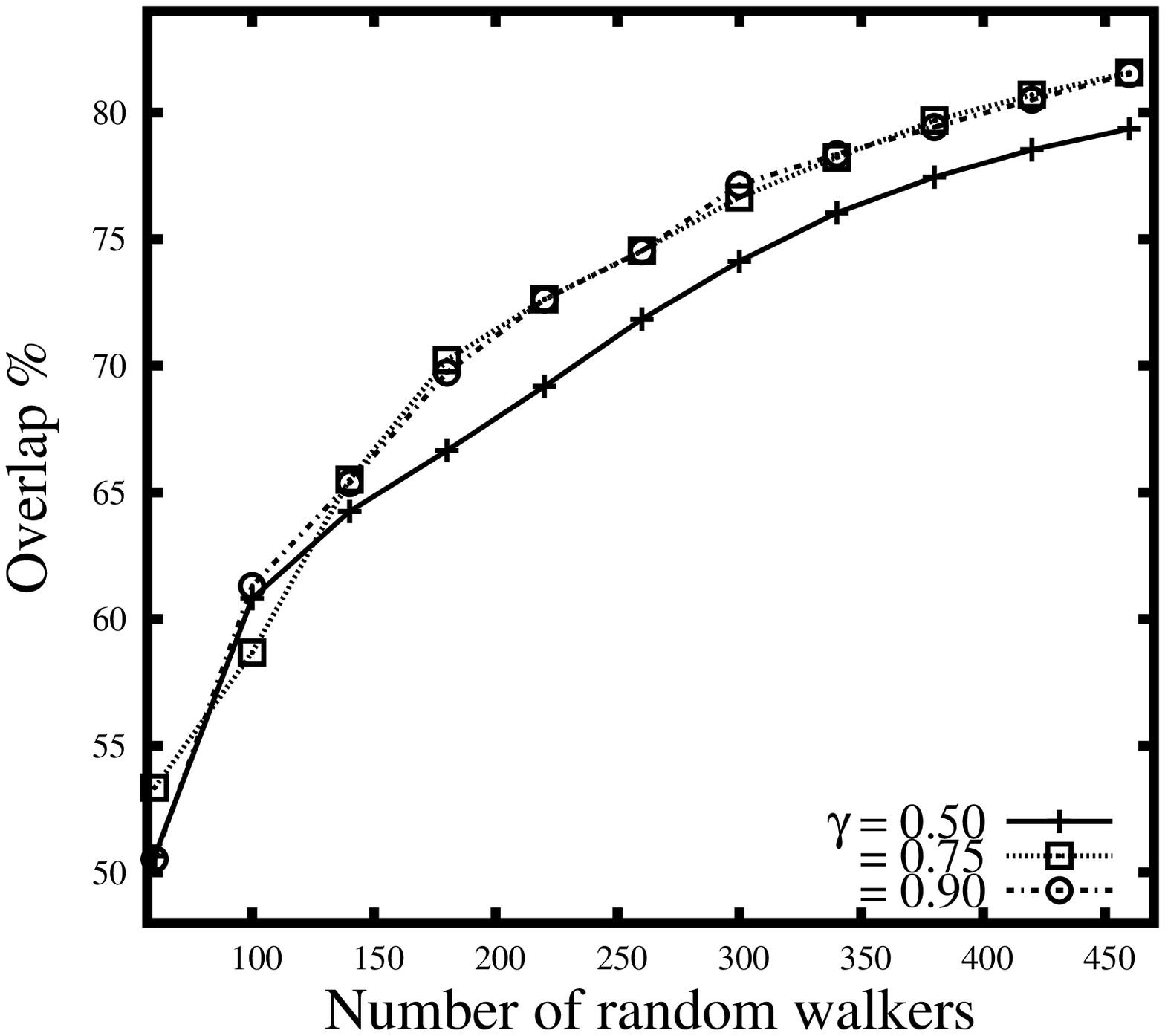}}
\subfigure[]{\includegraphics[width=2in,scale=1.0]{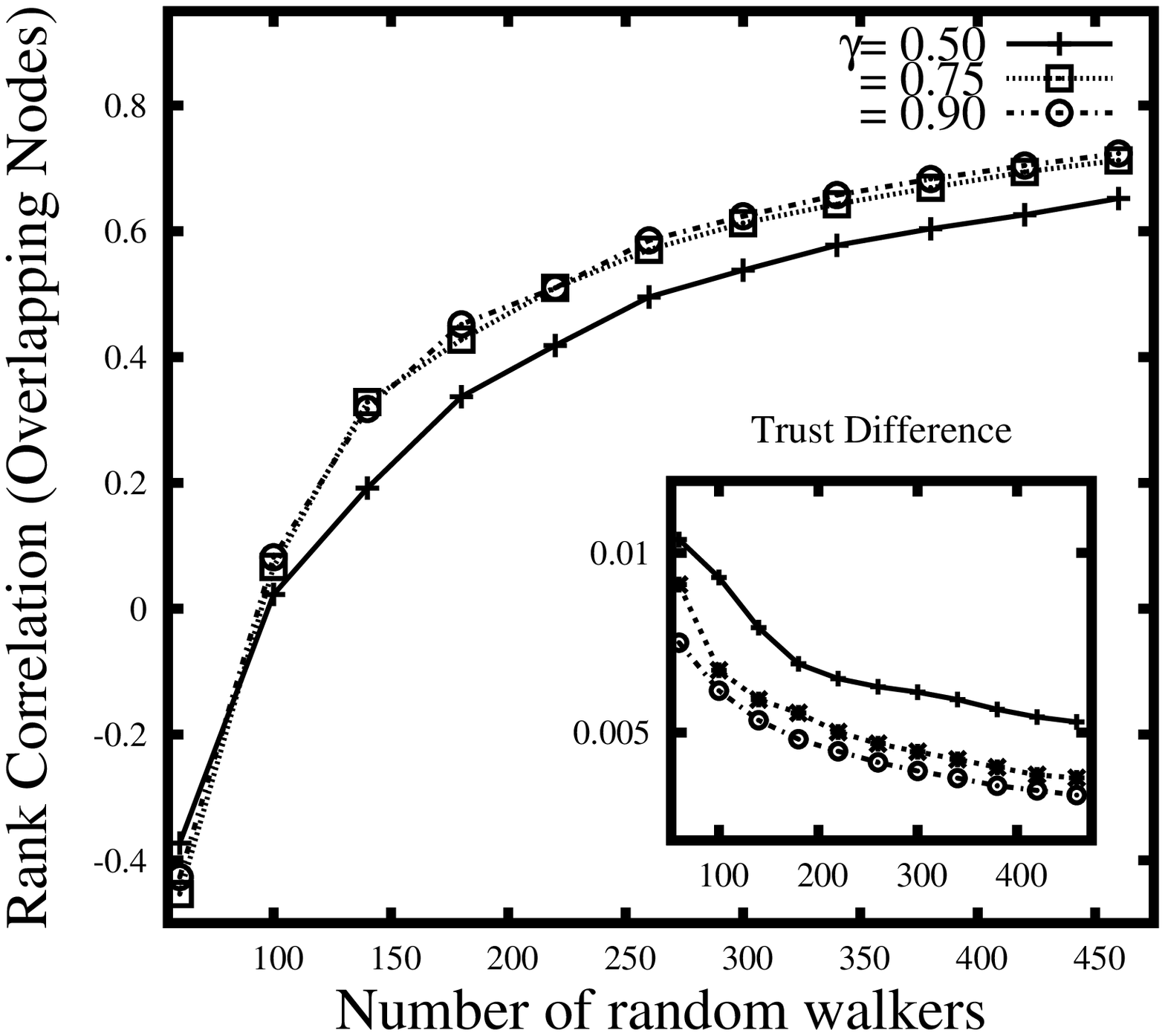}}
\subfigure[]{\includegraphics[width=2in,scale=1.0]{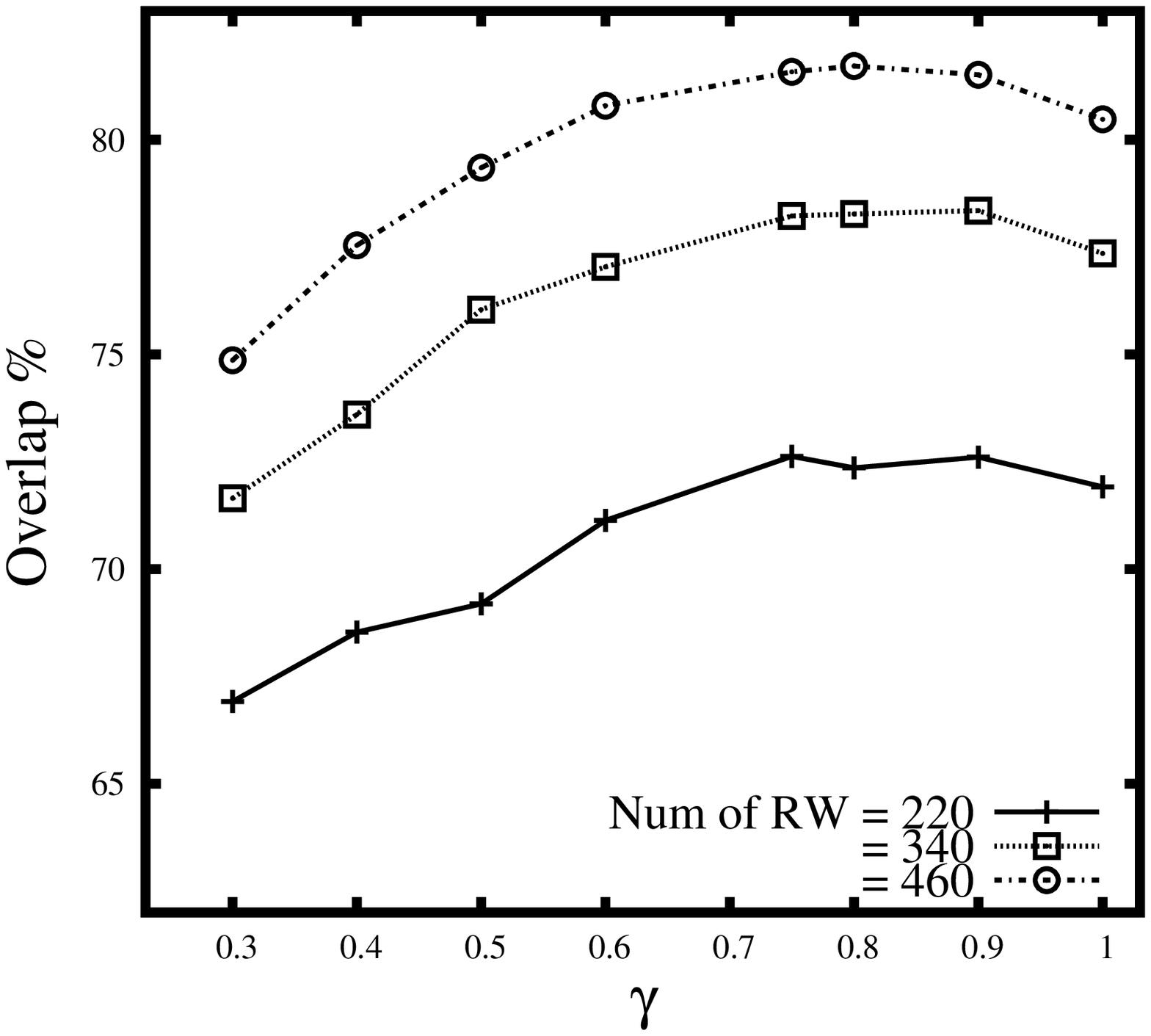}}
\subfigure[]{\includegraphics[width=2in,scale=1.0]{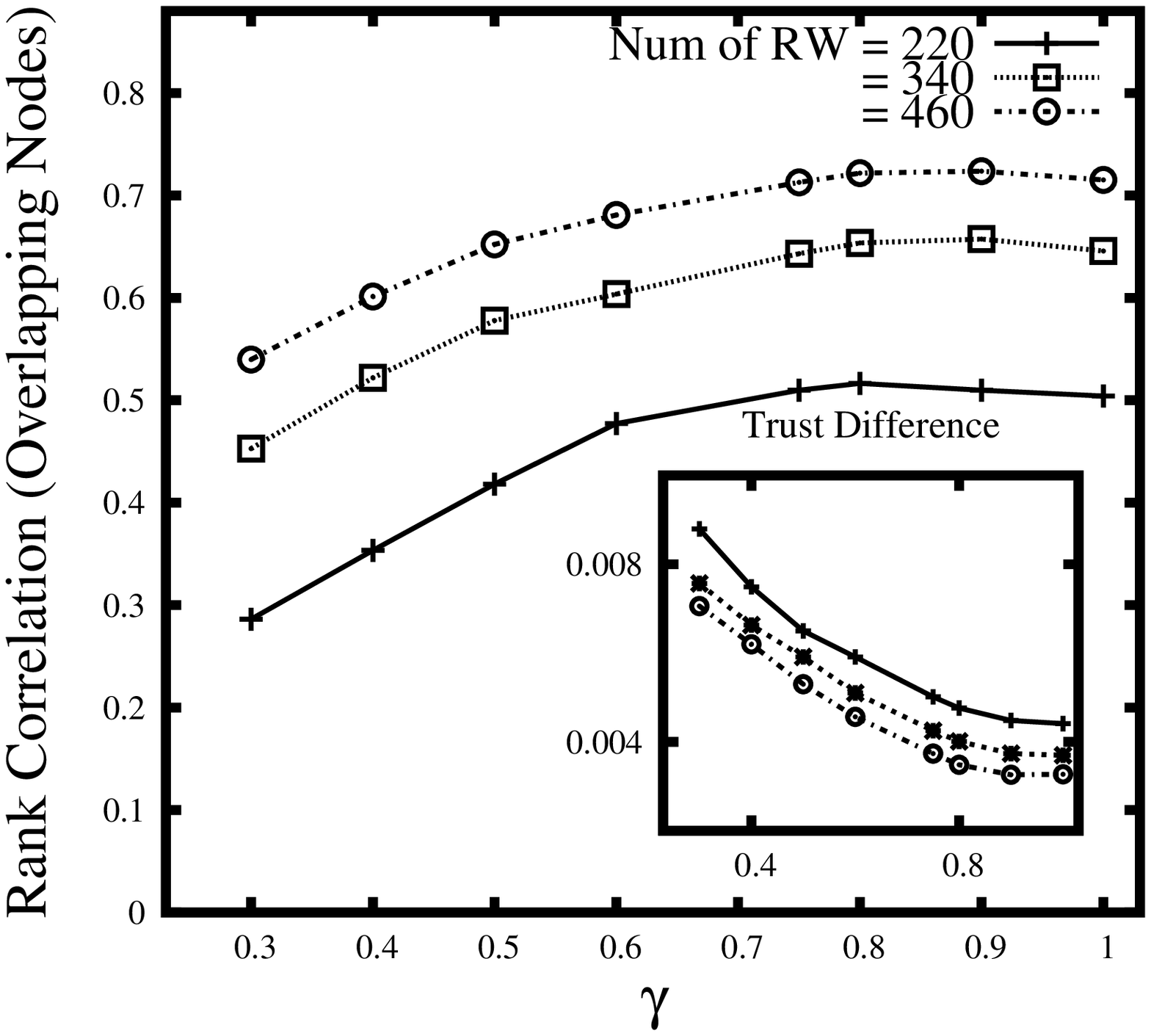}}
\caption{ \footnotesize \ref{fig:overlap}(a) -- \ref{fig:overlap}(d) show the simulation results for scale-free networks and 
\ref{fig:overlap}(e) -- \ref{fig:overlap}(h)
show the same results for Erd\"{o}s-R\'{e}nyi networks. 
\ref{fig:overlap}(a) and \ref{fig:overlap}(e) show the percentage overlap of the
top 5\% most trusted peers for TrustWebRank and our proposed algorithm for scale-free and Erd\"{o}s-R\'{e}nyi networks, respectively, for varying number of random
walkers. 
\ref{fig:overlap}(c) and \ref{fig:overlap}(g) show the same results for scale-free and Erd\"{o}s-R\'{e}nyi networks, respectively, for various values of $\gamma$. 
\ref{fig:overlap}(b) and \ref{fig:overlap}(f) show the rank correlation of the overlapping peers for  scale-free and Erd\"{o}s-R\'{e}nyi networks, respectively, and the figures in inset shows the
trust difference of the non-overlapping peers, for varying number of random walkers. \ref{fig:overlap}(d) and \ref{fig:overlap}(h) show the same result for scale-free and Erd\"{o}s-R\'{e}nyi networks, respectively, for various values of $\gamma$.
The networks considered are of 1000 nodes
with an average degree of 10. The value of $\beta$ considered for the TrustWebRank mechanism is 0.75.}
\label{fig:overlap}
\end{figure*} 

\section{Performance Analysis}\label{s:analysis}
In this section we critically analyze the proposed algorithm in terms of several metrics 
related to its correctness and efficiency. The initial analysis is based on the ability of the algorithm to identify the highly trusted
nodes in the network with respect to a particular user and then evaluating the minimum overhead required to identify a critical fraction of a set of trusted nodes. We then analyze the total coverage achieved
in terms of reaching a node by the random walkers and the efficiency achieved in terms of reducing
message overhead.  We observe the effect of the tuning parameters like the number of random walkers and
the dampening factor $\gamma$ on these metrics and compare the results with that of the TrustWebRank
mechanism. We finally extend the concept of personalization to evaluate a global perspective of the
nodes in a network and then analyze the efficiency of the algorithm in identification of the globally trusted peers in the network.
\subsubsection*{Experimental Setup}
Our analysis of the accuracy and efficiency of the algorithm is done on two types of networks, 
scale-free and Erd\"{o}s-R\'{e}nyi networks comprising of 1000 nodes with an average degree of 10.
The scale-free network is generated using the Barabasi-Albert preferential attachment model \cite{barabasi-science99}. The number
of random walkers in the simulations is varied from 60 to 460, whereas the value of $\gamma$ is
varied from 0.3 to 1.0. The value of $\beta$ is fixed to 0.75. This is as $\beta=0.75$ to 0.85 has been declared to be the best possible set of values by the proponents of TrustWebRank in \cite{walter-recsys09}.
We next discuss the performance of the algorithm for the above stated metrics. 
\subsection{Identification of Trusted Nodes}
One of the major motivations of the algorithm is to sample the trusted nodes in the network with 
respect to an individual node. To test the efficiency of the algorithm with respect to this objective,
we observe the ability of the algorithm to identify the \emph{top $p$ percentage of the most trusted
nodes}. We formally define this metric as follows:
\newtheorem{mydef}{Definition}
\begin{mydef}
 If $\mathcal{S}^{p}_H(i)$ and $\mathcal{S}^{p}_S(i)$ denote the set of top $p$ percentage of the most trusted nodes of node $i$, obtained
using our proposed algorithm and the TrustWebRank respectively, then the fraction of overlap of these two sets, given by 
$\frac{\mathcal{S}^{p}_H(i)\cap \mathcal{S}^{p}_S(i)}{|\mathcal{S}^{p}_H(i)|}$, provides a measure of efficiency
of the proposed algorithm in identifying the highly trusted nodes in network by an individual node.
\end{mydef}
\par We initially analyze the average fraction of overlap of the node set of top 5\% of the most trusted nodes for both TrustWebRank and our 
proposed mechanism for varying number of random walkers and for various values of $\gamma$. We also derive the rank correlation, $\mathcal{R}$,
of the overlapping
set of nodes,  using the Spearman rank correlation coefficient, and the absolute trust difference of the non overlapping nodes.
As stated earlier, the fraction of overlap provides a measure
of correctness of the proposed algorithm in identifying the most trusted nodes by an individual node. The rank correlation 
$\mathcal{R}$  provides an indication that if for a node $i$,  the set of its trusted nodes obtained using 
TrustWebRank can be sequenced 
in order of their trust values, then whether our algorithm can also rank the nodes in the same sequence. For the
non-overlapping nodes, the difference in the normalized trust values (that we term as \emph{Trust Difference} of the non-overlapping peers) 
is calculated to provide an indication of how close our algorithm has been
in missing the trusted node. We discuss the effect of the number of random walkers 
and the dampening factor $\gamma$ in this context.
\subsubsection*{Effect of the number of random walkers}
We initially discuss the effects of the number of random walkers on the above described parameter.
Simulation results in figures \ref{fig:overlap}(a) -- \ref{fig:overlap}(d) show the results for scale-free networks and the figures
in \ref{fig:overlap}(e) -- \ref{fig:overlap}(h) show the same for Erd\"{o}s-R\'{e}nyi networks, for increasing number of random walkers. 
Figures \ref{fig:overlap}(a) and \ref{fig:overlap}(e) show the average percentage overlap of the top 5\% most trusted peers for TrustWebRank and our
algorithm for scale-free and  Erd\"{o}s-R\'{e}nyi networks respectively. The percentage 
overlap for Erd\"{o}s-R\'{e}nyi networks is higher as compared to the scale-free networks for low 
value of $\gamma$ (0.50 in this case). This is due to the presence of a small number of hubs and large heterogeneity in the degree of nodes in scale-free networks; for low values of $\gamma$, the random walkers initially move towards the very few hub nodes and die out without exploring further trusted nodes. However, as can be observed that both the networks show similar trends, where
with increasing number of random walkers, the percentage of overlap increases and is nearly 80\% when the number of random walkers is
460, for higher values of $\gamma$. Further as shown in figures \ref{fig:overlap}(b) and \ref{fig:overlap}(f), the rank correlation of the
overlapping set of nodes is nearly 0.8 for both scale-free and  Erd\"{o}s-R\'{e}nyi networks indicating that the ordering of the nodes
in terms of their trust values as observed in TrustWebRank is highly maintained in our algorithm. The figures in inset shows the trust
differences of the non-overlapping nodes. It can be observed that although the differences are very high (nearly 0.01) for very
low number of random walkers but it decreases rapidly to as low as 0.001, when the number of random walkers is increased to 460. This 
indicates that using our algorithm, nodes are able to identify the most trusted nodes in the network with a very high probability. 
\subsubsection*{Effect of the Damping Factor $\gamma$}
Simulation results for certain random walkers shown in figures \ref{fig:overlap}(c) and \ref{fig:overlap}(g) indicates that 
the percentage overlap increases very rapidly with increasing values of $\gamma$, reaches a maximum point after which it starts decreasing. For
$\gamma=0.8$, the percentage overlap is nearly 80\% for both scale-free and Erd\"{o}s-R\'{e}nyi networks for 460 number of random walkers. Beyond a threshold value of $\gamma$, the random walkers tend to forget their respective starting source node and the number of
hits to a node become proportional to its degree. Thus the number of hits to a node gets biased by its degree and hence the percentage of overlap decreases.
This fact causes similar trends for the rank correlation of the overlapping nodes and the trust differences for the non-overlapping nodes,
as shown in figures \ref{fig:overlap}(d) and \ref{fig:overlap}(h).
\par Thus we observe that the damping factor $\gamma$ plays
an important role in our algorithm in determining the efficiency to predict the top trusted nodes in
the network. By selecting an optimal value of $\gamma$ (nearly 0.75-0.8 in this case) the algorithm can be highly efficient in 
identifying the trusted nodes in the network. 
\par We next observe the variation of the minimum number of random walkers required in recovering a certain number of trusted nodes in the network.

\begin{figure}[htbp]
\centering
\subfigure[]{\includegraphics[width=2.5in,scale=1.0]{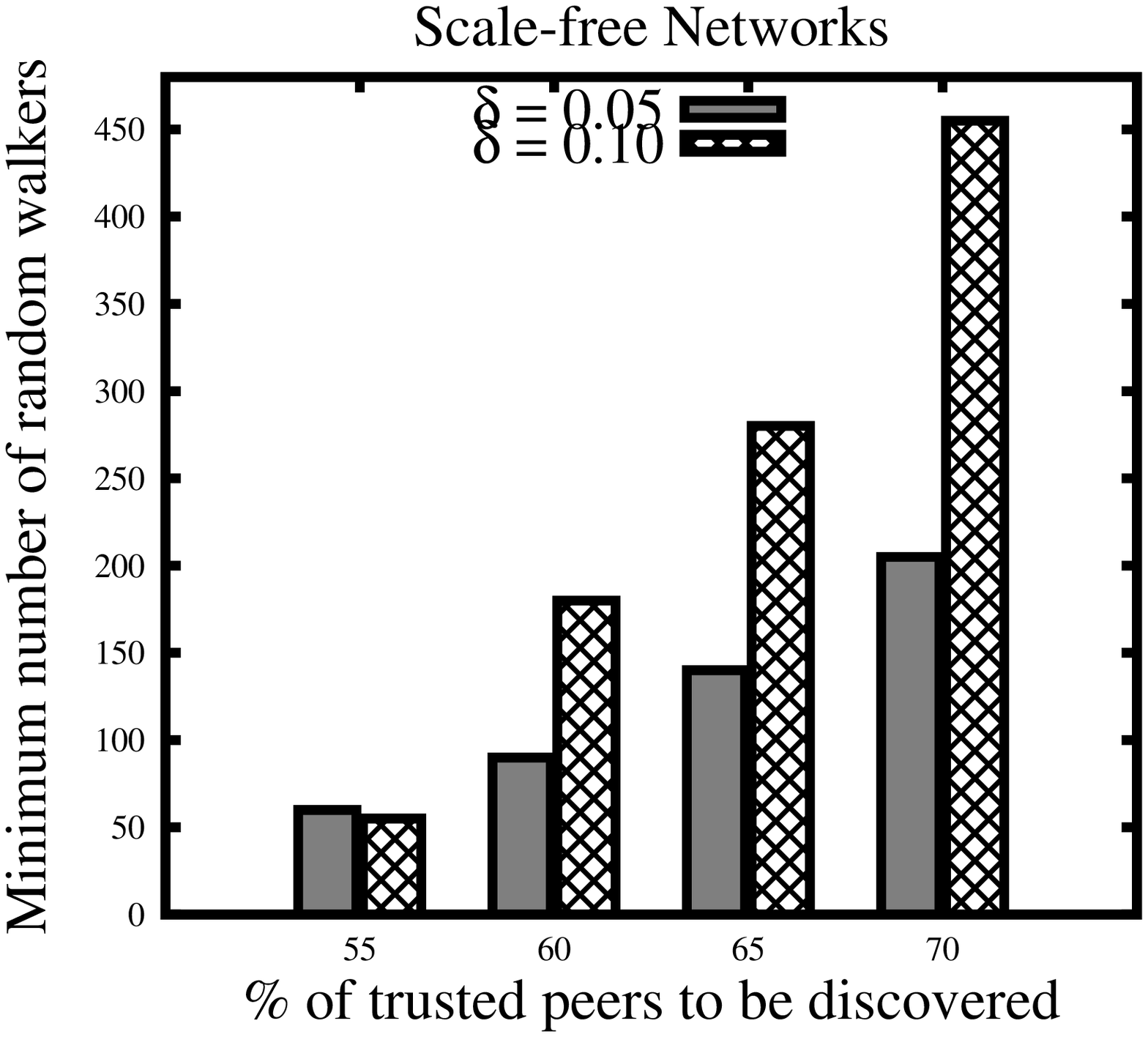}}
\subfigure[]{\includegraphics[width=2.5in,scale=1.0]{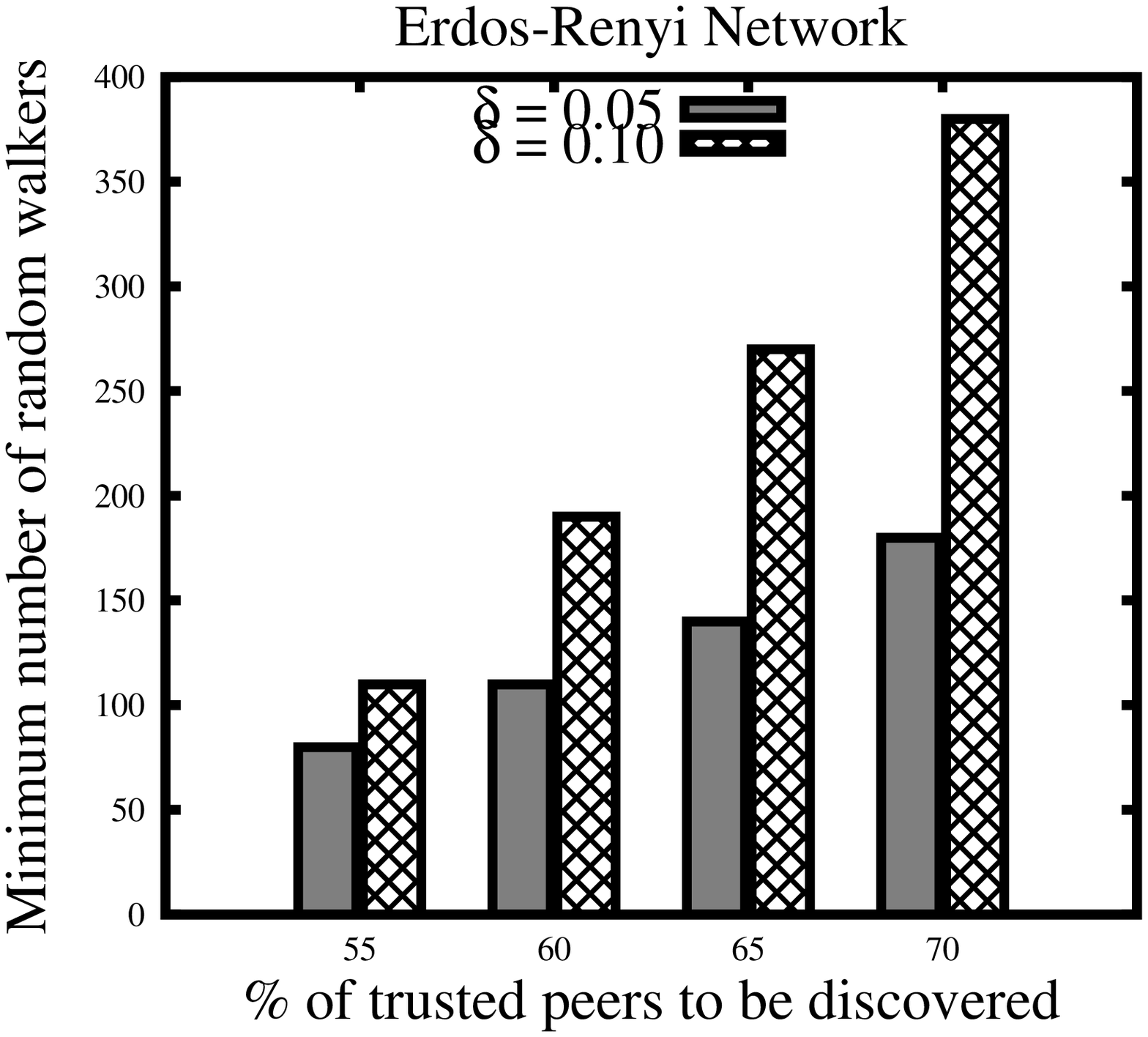}}
  \caption{  \footnotesize  \ref{fig:recoverability}(a) shows the minimum number of random walkers required to identify a certain percentage of the nodes in a set E formed from taking the top 5\% and 10\% ($\delta$ = 0.05 and 0.10 respectively) most trusted nodes in the network, for scale-free networks. \ref{fig:recoverability} shows the same for Erd\"{o}-R\'{e}nyi networks. The 
number of nodes used in the simulation is 1000 and the average degree is 10 for both these networks.}
\label{fig:recoverability}
\end{figure} 

\newpage
\subsection{Recoverability of Trusted Nodes}
In this section we observe how the minimum number of random walkers can be tuned to recover a certain fraction
of the trusted nodes. In various applications it may be so required that one needs a subset of trusted nodes. 
Let E denote a set of trusted nodes formed from selecting $\delta$ fraction of top-trusted nodes. 
Then the efficacy of the algorithm lies in discovering x\% of E with minimum number of walkers. 
\par As shown in figures \ref{fig:recoverability}(a) and \ref{fig:recoverability}(b) for scale-free and Erd\"{o}s-R\'{e}nyi network respectively, 
the minimum required number of random walkers is very low when one requires just 50\% of trusted nodes. Note that the value is roughly same 
for $\delta$ = 0.05, 0.10, although in essence the random walkers are discovering double the nodes in the second case. 
However, the number of walkers increases rapidly when one wants to obtain a higher fraction.  
The value increases even further when the cardinality of E ($\delta$) increases. This confirms that for identifying a very 
high fraction of trusted nodes in the network, the minimum number of
random walkers required will be enormously high; however, for smaller values of $\delta$, by appropriately tuning the
number of random walkers, a certain fraction of the top trusted nodes can be easily identified by using much less number of random walkers.  

\begin{figure*}[htbp]
\centering
\subfigure[]{\includegraphics[width=2.5in,scale=1.0]{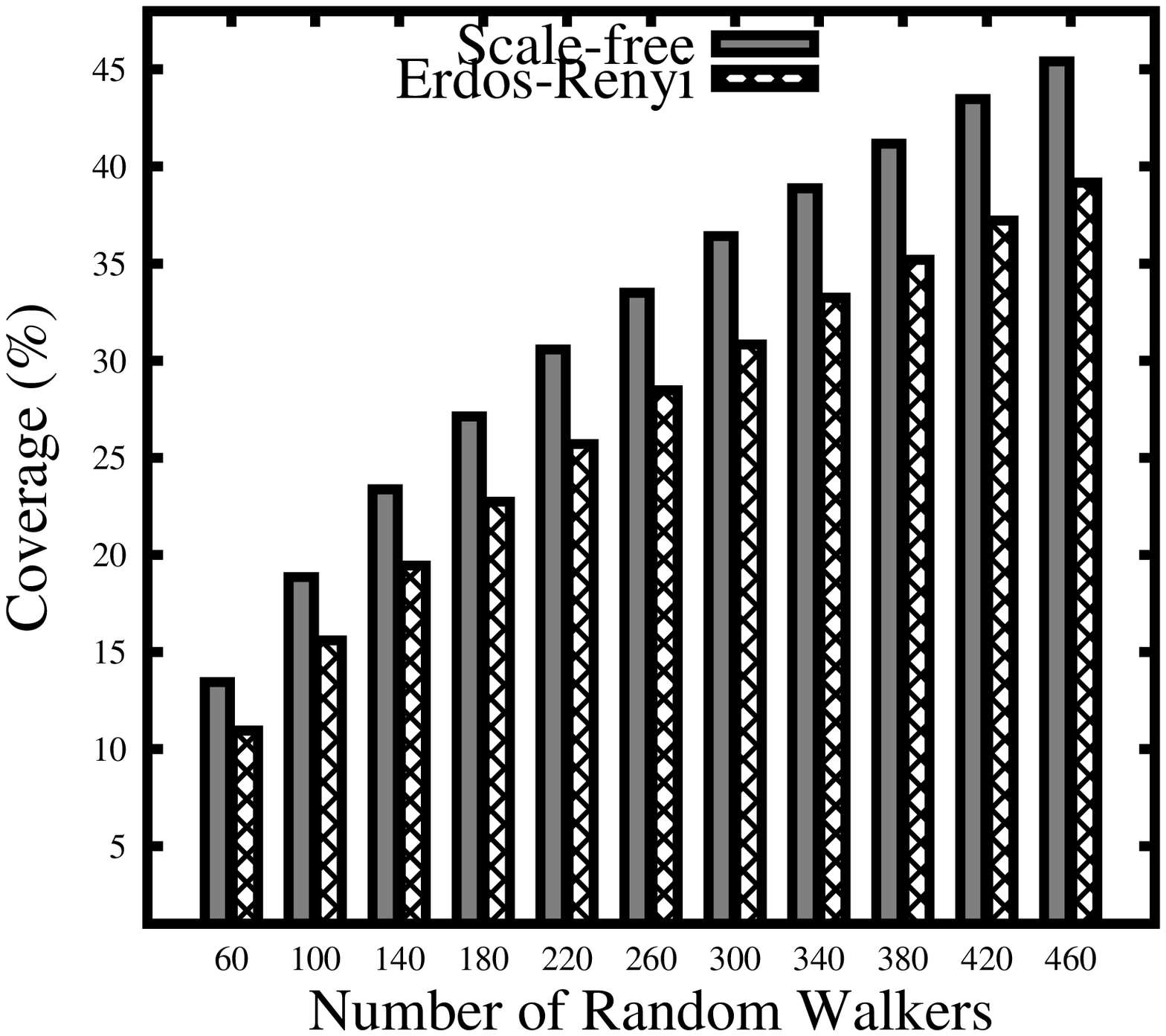}}
\subfigure[]{\includegraphics[width=2.5in,scale=1.0]{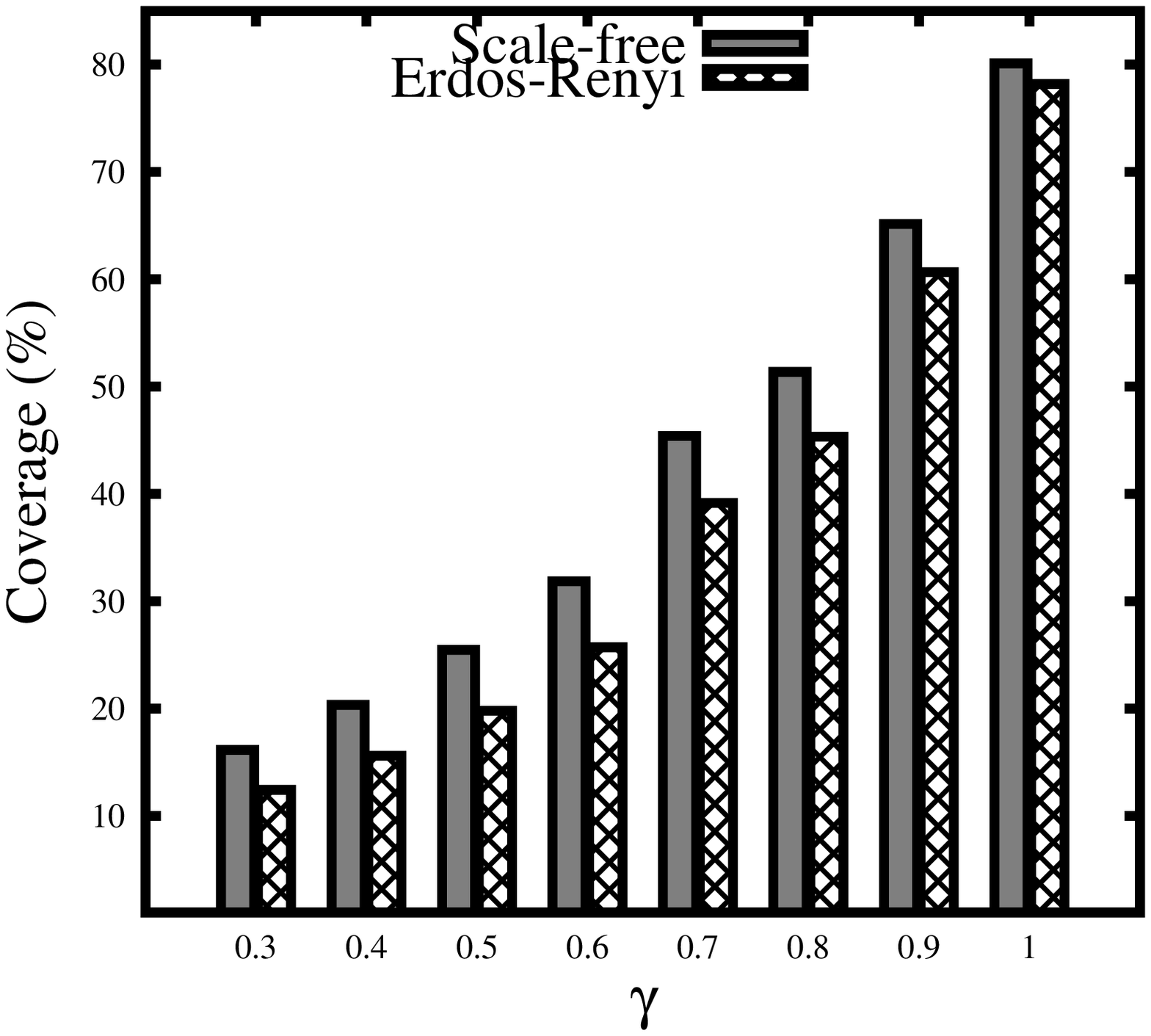}}
\subfigure[]{\includegraphics[width=2.5in,scale=1.0]{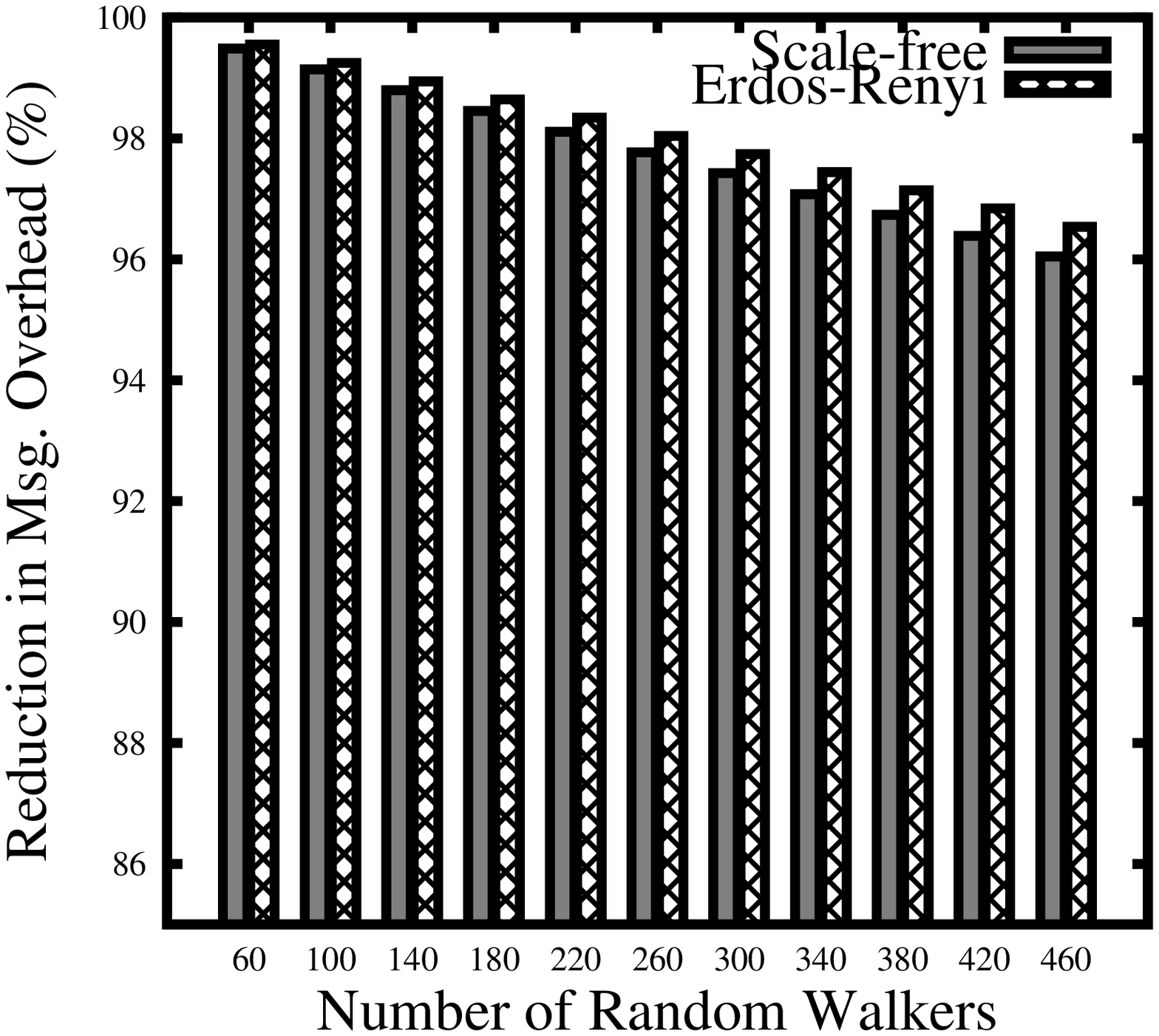}}
\subfigure[]{\includegraphics[width=2.5in,scale=1.0]{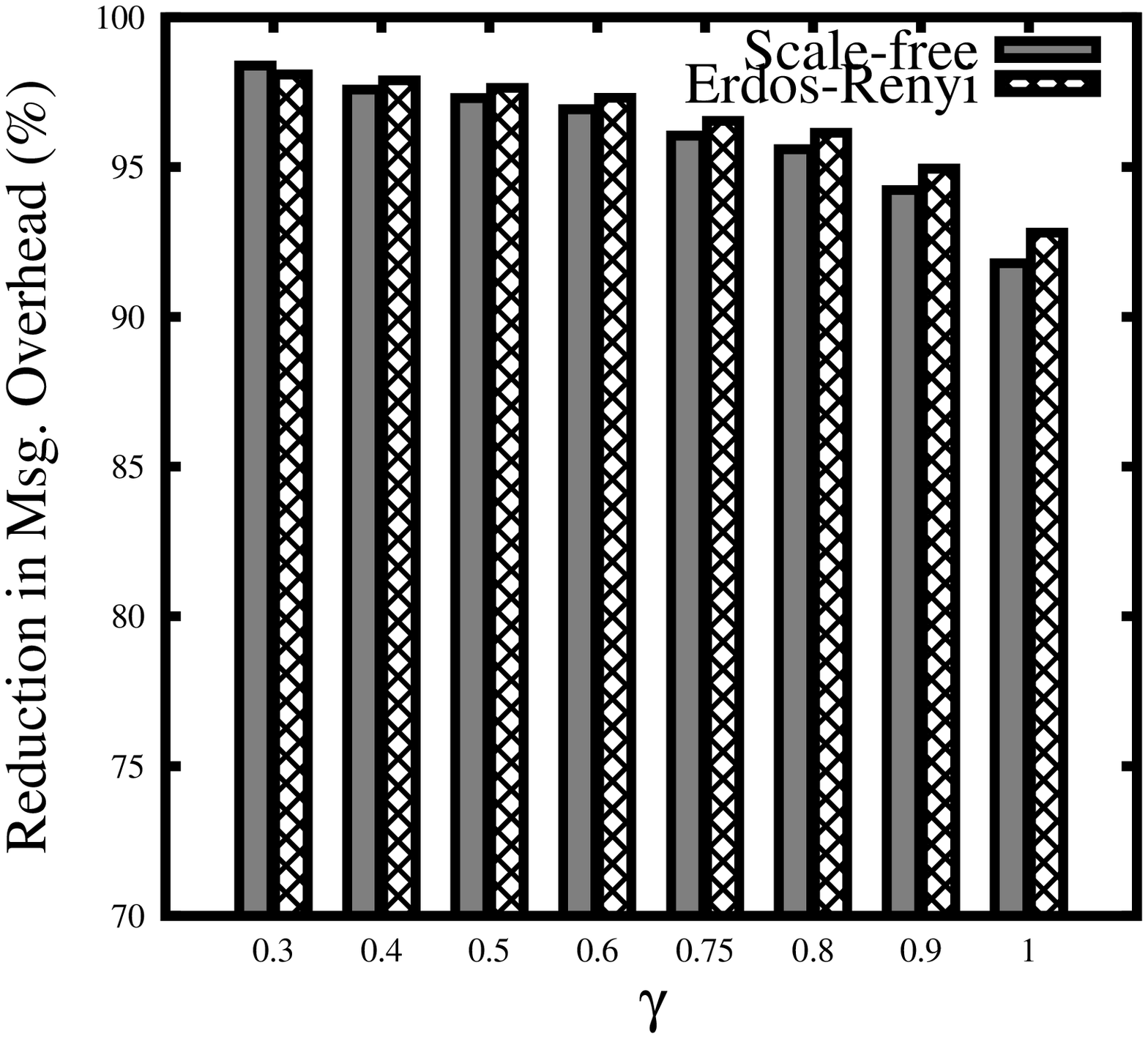}}
  \caption{ \footnotesize \ref{fig:msgoverhead-coverage}(a)and \ref{fig:msgoverhead-coverage}(b) compare the average coverage of the nodes for various random walkers and various values of $\gamma$, respectively, expressed in terms of the percentage of
nodes with non-zero trust values per node for both scale-free and Erd\"{o}s-R\'{e}nyi networks.  
\ref{fig:msgoverhead-coverage}(c) and \ref{fig:msgoverhead-coverage}(d) show the percentage reduction in message redundancy by using the proposed algorithm
for increasing number of random walkers and for various values of $\gamma$, respectively, as compared to the TrustWebRank mechanism for both scale-free and Erd\"{o}s-R\'{e}nyi 
networks.
The number of nodes in each of the simulation results is 1000 
and the average degree of the nodes is 10. The number of random walkers is varied from 60 to 460 and the value of both $\beta$ and $\gamma$
is considerd as 0.75.}
\label{fig:msgoverhead-coverage}
\end{figure*}

\subsection{Network Coverage}
We now analyze the network coverage of the peers achieved in terms of the number of unique nodes
explored by the random walkers from a source node and hence has an estimate of their trust values. We observe the network coverage for both the number of random walkers as well as the damping factor $\gamma$.
\subsubsection*{Effect of the number of random walkers}
We simulated the coverage of the nodes, i.e. the number of nodes in the network that has been reached by the random walkers 
and hence has a non-zero trust value. The results shown in figure \ref{fig:msgoverhead-coverage}(a) for $\gamma=0.75$ indicate that the coverage in case of scale-free network is higher as compared to Erd\"{o}s-R\'{e}nyi networks is higher; this is due to the lower diameter of the scale-free network as compared to the latter. However, in both cases, although the coverage
of the nodes are small for low number of walkers, but the coverage of the nodes increases rapidly with small increase in 
the number of walkers. The coverage of the nodes is as high as 45\% for the scale-free networks, when the number of walkers is 460. However for higher values of random walkers, the rate of increase of the coverage is slower for a fixed value of $\gamma$. 
\subsubsection*{Effect of the damping factor} 
We also simulated the coverage of the nodes for various values of $\gamma$ when the number of random
walkers is equal to 460. The simulation results shown in \ref{fig:msgoverhead-coverage}(b) shows that the damping factor has a huge effect on the coverage of the peers. Although the coverage of the peers for lower values of $\gamma$
is very low but with increasing values of the damping factor the coverage of the
nodes increase rapidly and reaches almost 80\% for a $\gamma$ of 1.0. 
Thus from figures \ref{fig:msgoverhead-coverage}(a) and \ref{fig:msgoverhead-coverage}(b), we conclude
that the coverage of the nodes can be tuned by a suitable controlling the number of random walkers
and $\gamma$.  Note that although we are sampling just 45\% of the nodes, we are identifying almost all
the most trusted nodes in the network.
\par We next analyze the efficiency of the algorithm in terms of the involved message overhead.
\subsection{Message Overhead}
In this section we discuss about the reduction in message overhead using our algorithm as compared to 
the distributed implementation of TrustWebRank in \cite{carchiolo-sci10}. 
We assume that a single unit of message is passed when a node communicates with other nodes for both
the TrustWebRank as well as our proposed algorithm. We observe the effect of both the number of
random walkers as well as the damping factor on the percentage reduction of number of messages passed by a node. 
\subsubsection*{Effect of the number of random walkers}
We simulated the reduction in the number of messages with increasing number of random walkers when
the value of $\gamma$ is 0.75. The simulation results shown in figure \ref{fig:msgoverhead-coverage}(c) indicate that for both scale-free and Erd\"{o}s-R\'{e}nyi networks, the proposed algorithm
sends nearly 96\% less number of messages as compared to the TrustWebRank, when the number of random
walkers is 460. The overhead involved is extremely low due to the fact that by using a constant number
of random walkers only a very small fraction of the edges are traversed.
\subsubsection*{Effect of the damping factor}
We also simulated the effect of $\gamma$ on the reduction of the number of messages when the number
of random walkers is equal to 460. Simulation results for scale-free network shown in figure \ref{fig:msgoverhead-coverage}(d) indicate that by increasing
the value of $\gamma$ from 0.3 to 1.0 leads to an increase in the message overhead by around 62\% (not shown in figure)
but the reduction in message overhead as compared to TrustWebRank is still around 92\%. 
\par We next attempt to evaluate a global trust measure of a node, in light of how trusted it is perceived  by other users in the 
network and observe whether the proposed mechanism can capture such a trust measure. 
\begin{figure}[htbp]
\centering
\subfigure[]{\includegraphics[width=2.5in,scale=1.0]{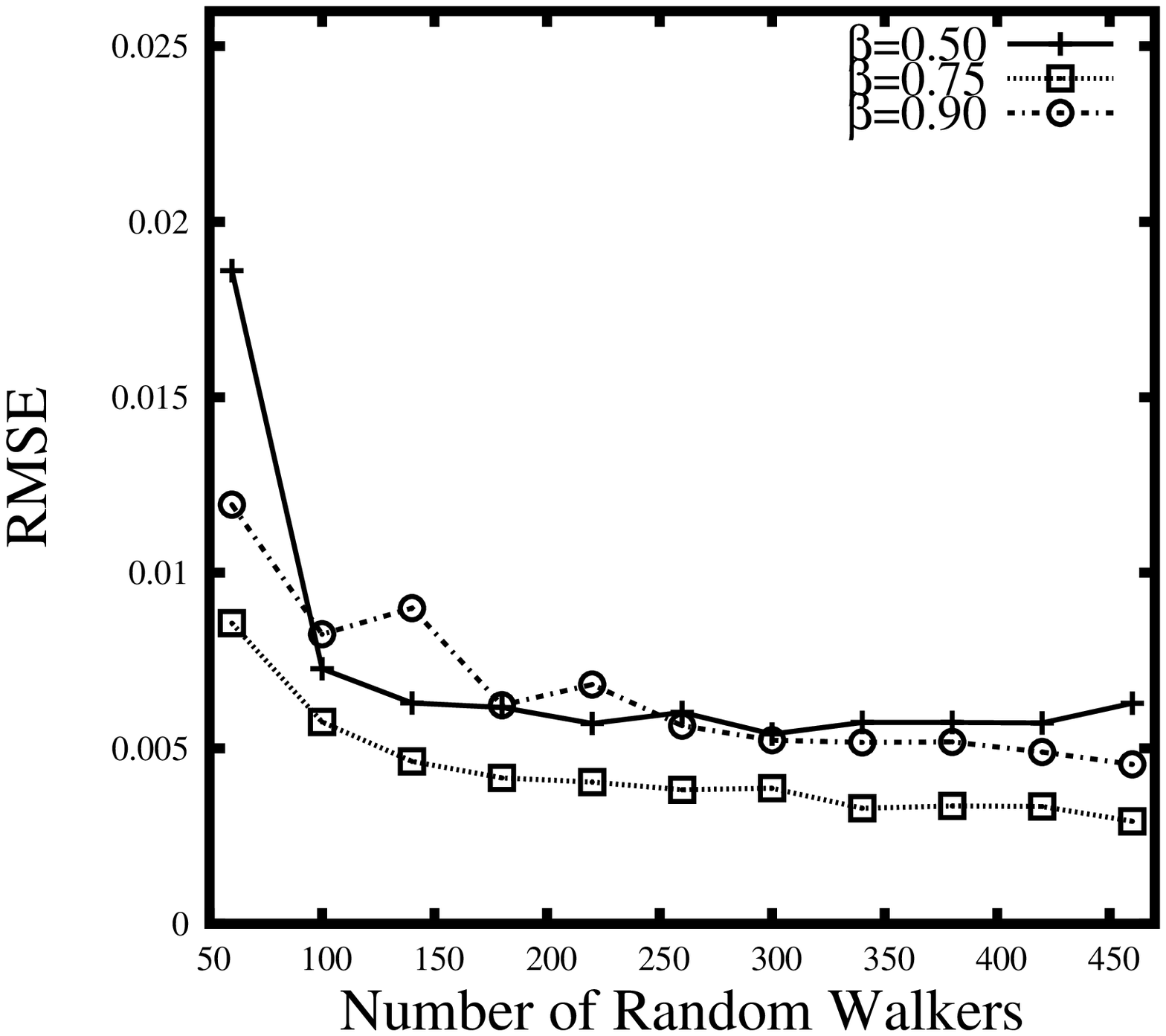}}
\subfigure[]{\includegraphics[width=2.5in,scale=1.0]{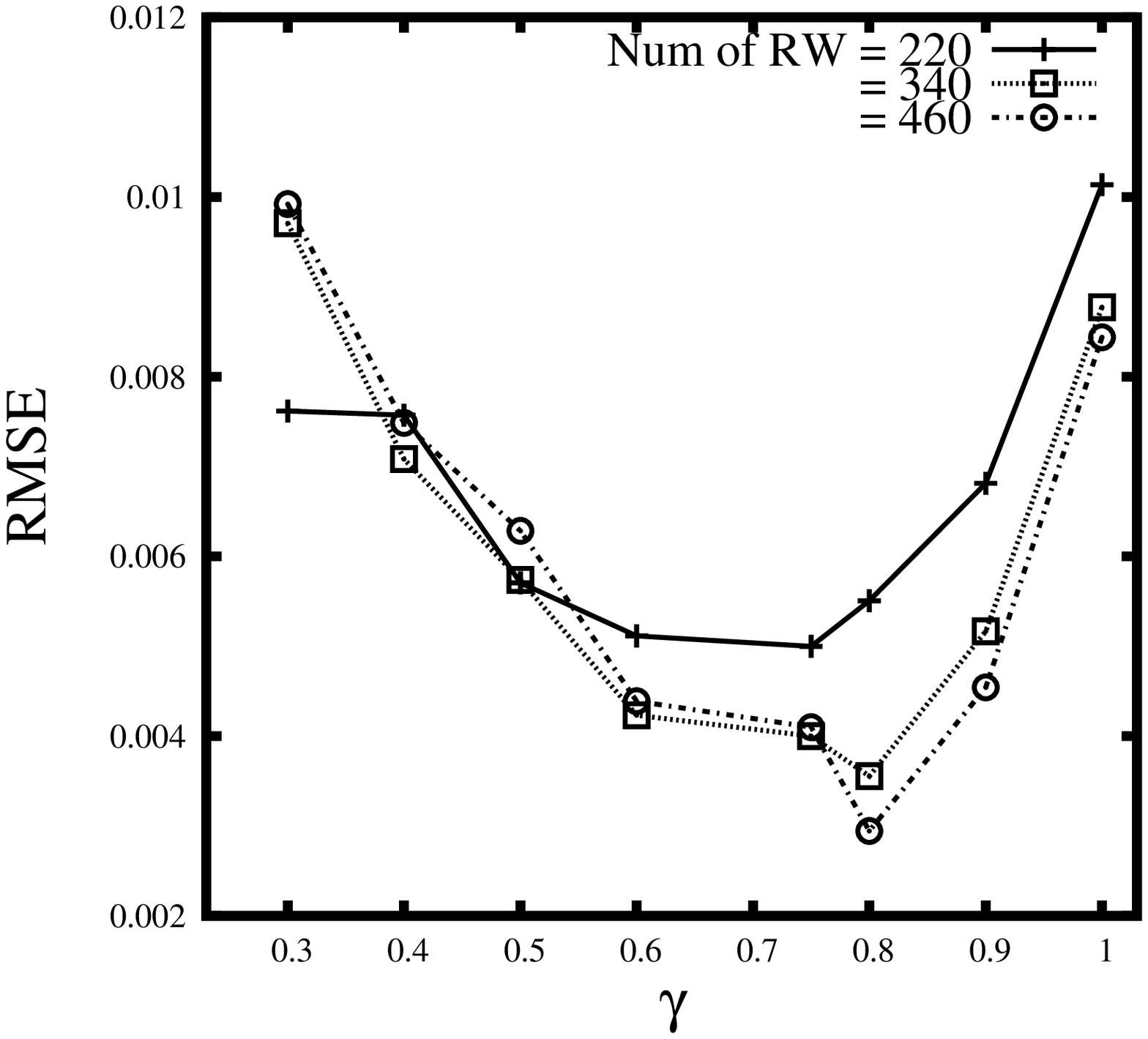}}
  \caption{ \footnotesize \ref{fig:RMSE-overhead}(a) and \ref{fig:RMSE-overhead}(b) show the simulation results for the RMSE values with respect to the number of random walkers and $\gamma$ respectively. The threshold value $\tau$ is considered to be 0.01. 
The networks considered are of 1000 nodes
with an average degree of 10. The value of $\beta$ considered for the TrustWebRank mechanism is 0.75}
\label{fig:RMSE-overhead}
\end{figure} 

\newpage
\section{From Local to Global Trust}\label{s:global}
Personalized trust values can in turn be used to calculate the global trust of a node, $i$, by aggregating all the personal trust values of other nodes in node $i$. Such a global trust can be helpful 
for a node in selecting initial trusted users in a community with whom the node had no prior interaction.
To capture the global trust of the nodes, we observe the \emph{global importance
of the nodes} in the network that we define as follows:
\begin{mydef}
We define the global importance, $I_j$, of a node, $j$, in the whole network as the fraction of the nodes in the network that have
normalized trust value in node $j$, greater than a given threshold $\tau$. Thus the global importance of a node gives a measure of what 
fraction of nodes in the network consider the node as trusted and have trust in it greater than the threshold $\tau$.
\end{mydef}
\par We observe the effectiveness of the proposed algorithm in identifying the globally trusted nodes in
the network. We analyze the root mean square error (RMSE) values of the global importance of all the nodes, for the TrustWebRank and our proposed mechanism. So if $I^{(T)}_i$ and $I^{(R)}_i$ denote the global importance values of node $i$, calculated using the TrustWebRank and our proposed mechanism, respectively, then the RMSE is given as $\sqrt{\frac{\sum_{i}(I^{(T)}_i-I^{(R)}_i)^2}{N}}$, where
$N$ is the number of nodes in the network. The threshold value $\tau$ in our simulations is considered as 0.01.
The RMSE values indicate whether the fraction of nodes that considers a node as trusted differs when the trust values are calculated using our mechanism and TrustWebRank.
\subsubsection*{Effect of the number of random walkers}
 Simulation results for scale-free networks, shown in figure \ref{fig:RMSE-overhead}(a) indicate that when the value of $\gamma$ is 
greater than 0.75, for moderately large 
number of random walkers, the RMSE values are very low --- around 0.006.  Since the global importance of a node 
indicates a fraction, a RMSE value of
0.006 indicates that the percentage of nodes that can be considered as trusted, when the trust values are calculated using TrustWebRank and 
our mechanism, differ by only 0.6\%, which is very low.
\subsubsection*{Effect of the damping factor}
The effect of $\gamma$ on the RMSE values of global importance of the nodes, for scale-free network, 
shown in figure \ref{fig:RMSE-overhead}(b), indicates that with increasing $\gamma$, nodes are able
to identify the globally trusted nodes more effectively. However beyond a threshold value of $\gamma$ (which is near to $\gamma=0.8$ in the present scenario),
 the RMSE values start increasing. The reason behind this trend is, as explained earlier, the coverage
of the nodes increases with increasing $\gamma$ thus leading to decreasing RMSE; however, with further
increase the number of hits at a node becomes biased by its 
degree leading to an increase in RMSE values.  
Thus for a suitable value of $\gamma$, the algorithm effectively identifies the globally trusted
peers in the network.

\section{Conclusion}\label{s:concl}
In this paper, we have proposed a random walk based mechanism to identify the trusted nodes in a network
with respect to a user in a distributed network.
We have shown that preferentially
sampling the nodes in the network based on their direct trust values identifies the highly trusted nodes
in the network, by exploring a very small subset of the total nodes and with much less overhead as 
compared to an exhaustive search. However the present analysis does not consider the dynamicity of
the nodes in the network. But it may be assumed that the trusted nodes in the network are relatively more stable as compared to other nodes. Hence these nodes can be easily located using our algorithm, as sampling in presence of such dynamicity will preferentially lead to the relatively more stable nodes. Further, this algorithm can be naturally extended to explore 
the non-transitive relations among the nodes (as discussed in \cite{guha-www04}) and compute trust based on these relations.

\bibliographystyle{plain}

\bibliography{paper}

\begin{thebibliography}{10}

\bibitem{barabasi-science99}
{Albert-L\'{a}szl\'{o}} {Barab\'{a}si} and {R\'{e}ka} Albert.
\newblock Emergence of scaling in random networks.
\newblock {\em Science}, 286(5439):509--512, Oct 1999.

\bibitem{bernard-2010}
Yvonne Bernard, Lukas Klejnowski, {J\"org} {H\"ahner}, and Christian
  {M\"uller-Schloer}.
\newblock Towards trust in desktop grid systems.
\newblock In {\em Cluster, Cloud and Grid Computing (CCGrid), 2010 10th
  IEEE/ACM International Conference on}, pages 637 --642, may 2010.

\bibitem{carchiolo-sci10}
V.~Carchiolo, A.~Longheu, M.~Malgeri, and G.~Mangioni.
\newblock A distributed algorithm for personalized trust evaluation in social
  networks.
\newblock In Mohammad Essaaidi, Michele Malgeri, and Costin Badica, editors,
  {\em Intelligent Distributed Computing IV}, volume 315 of {\em Studies in
  Computational Intelligence}, pages 99--108. Springer Berlin / Heidelberg,
  2010.

\bibitem{guha-www04}
R.~Guha, Ravi Kumar, Prabhakar Raghavan, and Andrew Tomkins.
\newblock Propagation of trust and distrust.
\newblock In {\em Proceedings of the 13th international conference on World
  Wide Web}, WWW '04, pages 403--412, New York, NY, USA, 2004. ACM.

\bibitem{karbhari-pam04}
Pradnya Karbhari, Mostafa Ammar, Amogh Dhamdhere, Himanshu Raj, George Riley,
  and Ellen Zegura.
\newblock Bootstrapping in gnutella: A measurement study.
\newblock In Chadi Barakat and Ian Pratt, editors, {\em Passive and Active
  Network Measurement}, volume 3015 of {\em Lecture Notes in Computer Science},
  pages 22--32. Springer Berlin / Heidelberg, 2004.

\bibitem{li-ithet10}
Na~Li, Sandy El~Helou, and Denis Gillet.
\newblock Trust-based rating prediction for recommendation in web 2.0
  collaborative learning social software.
\newblock In {\em Proceedings of the 9th international conference on
  Information technology based higher education and training}, ITHET'10, pages
  197--201, Piscataway, NJ, USA, 2010. IEEE Press.

\bibitem{liu-cikm09}
Xin Liu, Anwitaman Datta, Krzysztof Rzadca, and Ee-Peng Lim.
\newblock {StereoTrust: A group based personalized trust model}.
\newblock In {\em Proceedings of the 18th ACM Conference on Information and
  Knowledge Management}, CIKM '09, pages 7--16, New York, NY, USA, 2009. ACM.

\bibitem{massa-recsys07}
Paolo Massa and Paolo Avesani.
\newblock Trust-aware recommender systems.
\newblock In {\em Proceedings of the 2007 ACM conference on Recommender
  systems}, RecSys '07, pages 17--24, New York, NY, USA, 2007. ACM.

\bibitem{matsuo-www09}
Yutaka Matsuo and Hikaru Yamamoto.
\newblock Community gravity: measuring bidirectional effects by trust and
  rating on online social networks.
\newblock In {\em Proceedings of the 18th international conference on World
  wide web}, WWW '09, pages 751--760, New York, NY, USA, 2009. ACM.

\bibitem{walter-recsys09}
Frank~E. Walter, Stefano Battiston, and Frank Schweitzer.
\newblock Personalised and dynamic trust in social networks.
\newblock In {\em Proceedings of the 3rd ACM Conference on Recommender
  systems}, RecSys '09, pages 197--204, New York, NY, USA, 2009. ACM.

\bibitem{Wierzbicki-2010}
Adam Wierzbicki and Adam Wierzbicki.
\newblock Trust management.
\newblock In {\em Trust and Fairness in Open, Distributed Systems}, volume 298
  of {\em Studies in Computational Intelligence}, pages 71--143. Springer
  Berlin / Heidelberg, 2010.

\bibitem{zhao-icccn09}
Huanyu Zhao and Xiaolin Li.
\newblock {VectorTrust: Trust vector aggregation scheme for trust management in
  peer-to-peer networks}.
\newblock In {\em Proceedings of 18th Internatonal Conference on Computer
  Communications and Networks, 2009}, ICCCN '09, pages 1 --6, Aug. 2009.

\bibitem{zhou-ipdps06}
Runfang Zhou and Kai Hwang.
\newblock {Trust overlay networks for global reputation aggregation in P2P grid
  computing}.
\newblock In {\em Proceedings of 20th International Conference on Parallel and
  Distributed Processing Symposium}, IPDPS '06, page 10 pp., April 2006.

\end{thebibliography}

\end{document}